\documentstyle[11pt,paspconf]{article}
\input epsf.sty

\begin{document}

\title{Some results based on the EROS microlensing survey }

\author{J.P. Beaulieu$^*$}\footnote{* Member and on behalf of the EROS collaboration}
\affil{Kapteyn Institute, Groningen, The Netherlands}
\author{H.J.G.L.M. Lamers, W.J. de Wit}
\affil{SRON and Astronomical Institute, Utrecht, The Netherlands}



\begin{abstract}
We will review some results based on the EROS and the on going EROS-2
microlensing surveys to search for dark matter in the Galactic halo
via microlensing effects on LMC/SMC stars. Microlensing surveys
provide systematic observations for millions of 
stars  over a long period in both Magellanic
Clouds and give birth to very powerful database. We will quickly 
review the results obtained on pulsating stars and 
we will detail the similarity and differences of the
Herbig Ae/Be stars in the Magellanic Clouds compared to their 
Galactic counterpart.
\end{abstract}





\section{Observations made by the EROS microlensing survey }

EROS (Exp\'erience de Recherche d'Objets Sombres) is a French
collaboration between astronomers and particle physicists to search 
for baryonic dark matter
in the Galactic Halo by microlensing ( Paczy\`nski, 1986) on stars of 
the Magellanic
Clouds (MCs).  A compact object in the Galactic
Halo passing close enough to the line of sight to a background star 
in the Magellanic Clouds induces an increase in the apparent 
brightness 
of the star. This phenomenon
occurs owing to the alignment of the observer, the deflector and the
background star. Assuming a standard corotating halo the time scale 
$\tau_0$  of a
microlensing event is given by the relation $\tau_0 = 
70\sqrt{M/M_\odot} $ days
where $M$ is the mass of the deflector. In order to be
sensitive to a wide range of mass for compact objects 
($10^{-7}-10^{-1}M_\odot$)
two complementary approaches have been developed.

The first involved photographic monitoring of 6.4 million stars over 
a
 $5\times 5 $ degree field using the ESO Schmidt telescope from 1990
 to 1994. About 380 plates were taken
once a night if possible in two colours $B_J$
 and $R_C$. Isolated stars can be found down to magnitude 20 with
 typical photometric uncertainties of 0.3 mag.

The second approach was to be sensitive to low mass compact objects. 
Between 1991 and 1994 about 15000
CCD images were taken with two broad bandpass filters $B_E$ and $R_E$ 
of an area of 
1 x 0.4 degrees centered  on the bar of the LMC. Another field of 1 x 
0.4 degrees had been
chosen in the SMC between 1993-1995 and about 6000 images where 
taken. 
Lightcurves of 270 000 stars exist with up to 48 points per  night ! 

EROS have been upgraded recently and entered the EROS-2 phase.
EROS-2 uses the dedicated Marly 1m telescope at ESO La Silla. The 
prime
focus is equipped with a focal reducer and a dichroic beam splitter 
with 
a mosaic of eight CCD $2048 \times 2048$ in each channel. The total 
field
is $0.7 \times 1.4$ deg. 
The collection of data started in July 1996 for a minimal observing
period of 4 years. EROS-2 observes about 20 million stars a night, 
and
monitor the LMC, SMC, the Galactic Bulge and some fields in the 
spiral arms  (with approximately a daily sampling rate).

\section{Microlensing related results}

The first success of microlensing surveys have been to show that the
experiment was feasible. Indeed, the first microlensing events 
towards the LMC 
were announced almost simultaneously in October 1993 by the two
competing teams EROS (Aubourg et al., 1993) and MACHO (Alcock et al.,
1993 see Cook this volume). Now, about 18 have been observed toward 
the LMC and
the SMC. Despite the fact that some of the microlensing candidates 
are 
variable stars (EROS1 is a Be star, Beaulieu et al., 1995, EROS2 is 
an
A0-2 eclipsing binary, Ansari et al., 1995, EROS-97-SMC-1 
-also known as MACHO-97-SMC-1- is an eclipsing binary, Palanque et 
al., 1998), the vast
majority seems to be due to genuine microlensing. Moreover, 
we still do not know an intrinsic physical mechanism to mimic the 
observed 
photometric variation, and the microlensing interpretation is still
favored.

EROS and MACHO measured consistent with each other optical depth for 
gravitational lensing toward the LMC of 
 $\tau_{LMC} = 0.82^{+1.1}_{-0.5}~10^{-7}$ (Renault et al., 1997a)  
and 
 $\tau_{LMC} = 2.9^{+1.4}_{-0.9}~10^{-7}$ (Alcock et al., 1997) 
respectively.
A subsequent combined EROS MACHO study gave  $\tau_{LMC} = 
2.1^{+1.3}_{-0.8}~10^{-7}$ (Alcock et al., 1998).
The expected optical depth to gravitational lensing towards the LMC 
due to known stellar
populations is expected to be smaller than $0.5~10^{-7}$. 
The generally favored hypothesis is that all the observed 
microlensing are
due to lenses in the Halo. 
We should underline that in most cases, we do not have direct 
access to the different parameters of the  system. The parameters 
are degenerate, therefore we do not know where 
the lens is (in halo or in the Clouds). Based on likelihood analysis 
it is concluded that half of the dark halo of our 
Galaxy is composed of MACHOs with a mass of $\sim 0.5 M_\odot$. 
If they are old white dwarfs it would lead to various serious
problems (cooling rate of white dwarves, shape of the IMF for the
first generation of stars, star formation rate etc).
If not, what are they ? Do they belong to the halo ?

On theoretical ground, Zhao (1998) had suggested  the elegant idea of 
a tidal streamer front of the LMC 
as the cause of the microlensing signal. Zaritsky \& Lin (1997)
claimed that a vertical extension of the red clump observed in LMC
CMDs (VRC hereafter) was indeed the signature of such a streamer 
located at 35 kpc.  
Beaulieu \& Sackett (1998) confirmed the detection of the
VRC, but showed that it resulted from stellar evolution within the 
LMC, i.e. younger
clump stars. The observed colors and densities in the CMDs are
reproduced with a very simple model.   Moreover, a radial velocity
survey of VRC stars (Ibata, Lewis \& Beaulieu, 1998, ApJL submitted)
showed that they have the same systemic velocity as the LMC field 
stars, and the same velocity dispersion. The chance of a tidal streamer to have 
these kinematic properties are indeed low. The VRC is not the signature of 
a tidal streamer. However, it would be interesting to do an extensive 
search of tidal streamers very close to the LMC.

Recent studies towards the SMC by EROS (Palanque et al., 1998, Afonso
et al., 1998)  or by PLANET (Albrow et al., 1998),  (see Milsztajn this volume for a 
detailed  account on SMC microlensing) suggest that the events for which the
degeneracy between the parameters have been partially broken
are due to lensing in the clouds themselves, not to halo lenses.
The LMC and SMC self lensing may have been systematically 
underestimated.

\section{Variable star results}

The microlensing surveys generate very large homogenous database
of stars observed over a long period of time in the two Magellanic
Clouds. They provide gold mines for stellar studies (see also 
results from the MACHO collaboration Welch et al., Alves et al., this
volume). 
We will quickly review the results on pulsating variables, Nova and 
eclipsing binaries. Then, we will present a status report of our study on 
pre-main sequence stars in the Magellanic Clouds.
 
\subsection{Pulsating variables, Nova, Eclipsing Binaries}

We have searched systematically the EROS data base for variable stars
using the AoV method (Schwarzenberg-Czerny 1989).
We have built catalogues of variable stars in the LMC and in the SMC. 
We will give a short report of the results on variable stars by EROS. 
A differential study of about 550 Cepheids from LMC and SMC have been
conducted, to derive constraints on stellar pulsation theory (position of
resonance centers, and beat Cepheids) and to find a
subtle metallicity effect on the Period Luminosity relation
(Beaulieu et al., 1995, 1997a, 1997b, Sasselov et al., 1997).
A larger sample of Cepheids from LMC and SMC has been observed as
a pilot project of the EROS-2 survey (Bauer et al., 1998, Marquette,
Graff, Ripepi et al., this volume).
 
The first results on $\sim$2000 LMC RRLyraes from EROS
Schmidt plates has been presented by Hill \& Beaulieu (1997).
The period histograms of RRLyrae from the bar and from the field are
similar. From the field they confirm the histograms found in the
literature. The metallicity derived for these stars 
suggests that the old population in the field might have formed a
little  earlier than the oldest cluster in the LMC.
    
Grison et al., (1995) give a catalogue of 79 Eclipsing binaries in 
the bar of the LMC ranging from close to well-detached type. They can be
used for accurate distance determination of the LMC or as a test of 
stellar pulsation models.
 
de Laverny et al.,  (1998) report the observation of a slow Nova in 
the SMC. Thanks to the sampling rate of the observations (2 points/ 20 
min), low amplitude short time scale variation has been observed at 
maximum.
This nova is of DQ Her type, similar to Galactic novae of the same
class. Large Helium enhancement in the shell is found, and C and O
enrichments are suspected.

Beaulieu et al., (1997), Koll\`ath et al., (1997) discuss the 
properties and compute linear and hydro models of a particular variable star 
observed in  the bar of the LMC. It has 2 modes  of pulsation coupled by a 3:2
resonance,  shows a strong $H \alpha$ emission and has a spectral type
late O to early B. Its nature remains a puzzle. Is it a pre-main 
sequence star  or post-AGB ?

\subsection{Pre-main sequence stars in the Magellanic Clouds}

The process of star formation of low mass stars and their subsequent
evolution  to the main sequence in external galaxies is essentially
unknown. All the studies of star formation in external galaxies so 
far have been restricted to massive star clusters, such as the 30 Dor
region in the LMC. This is because massive young stars are very
luminous and the HII regions are signposts of recent massive star 
formation. The population of low mass stars can be studied in such 
cluster by deep photometry. In this way, the low mass population of
the 30 Dor and around SN1987A has been studied (see Panagia, Walborn,
Zinnecker and references in this volume). However, such
studies are always biased to conditions in massive  star clusters
which are not characteristic for the formation and the IMF of the
majority of the  less massive stars.

Using data from the EROS survey in the bar of the LMC, we discovered
a population of Herbig Ae Be stars (HAeBes hereafter) on the basis
of their characteristic irregular photometric variations and their 
colour.
Subsequent optical spectra confirmed the spectral types B, and the
presence of strong $H \alpha $ emission. The first seven have been
published by Beaulieu et al., (1996), Lamers et al., (1998), 
and the new 15 HAeBes by de Wit et al., (1999). The LMC HAeBes are
located in a large star forming region in the Bar with strong IR
excess  at IRAS wavelengths. Two stars with very similar
characteristics were found in the SMC in a pilot search using EROS-2 
data. These stars are the analogues of the Galactic pre-main sequence HAeBe 
stars.

The fields near the HAeBe stars are ideally suited for the study of 
the low mass star
formation, because they contain many T Tauri stars. In
fact, a study of these fields  at ESO in B,V,R,I and H$\alpha$ down 
to  $19.5^m$,  reveals colour magnitude diagrams that contain hundreds
of emission line  objects (HAeBe and T Tauri stars) down to the detection limit. 
Whereas few dozen emission lines objects are observed in the control fields
outside the star forming region.
The magnitudes of the optical study limits the determination of the
IMF to  $M_v \simeq +0.5^m$ or $M_* \simeq 3 M_\odot$. 

The location of the Galactic PMS in the HRD shows an upper envelope,
called "the birthline".  It marks the location of the stars at the 
end of the
fast accretion rate when they become optically visible through the
dust. The birthline has been predicted for different accretion rates
by Palla \& Stahler (1993) and Bernasconi \& Maeder (1996). 
There are  no Galactic HAeBe stars with $M_* > 10 M_\odot$, with the 
exception of a few stars which, however, have uncertain distances.
However five of the six HAeBe stars in the LMC,
studied in detail by Lamers et al. (1998)  
have luminosities above the observed and predicted
Galactic birthline by factors 3 to 10 (see figure 1) !
 Does the birthline depend on metallicity? If so, we should find
a systematic trend from SMC, LMC to Galaxy. What could be the reason?
 Is the accretion rate dependent on metallicity? Faster accretion 
would result in a higher birthline.
 Or does dust play a more crucial role than predicted in determining 
the luminosity of the star when its optical radiation emerges?
(A lower metallicity may result in a smaller dust/gas ratio).

Galactic HAeBe stars show irregular photometric variations with 
amplitudes of typically a few tenths of a mag. and occasionally a deep minimum up 
to 2  mag. Normally, Galactic HAeBe stars  
are "redder-when-fainter". This is explained by irregularity of the
surrounding dust clouds orbiting the star, and confirmed by
simultaneous photometric and polarimetric observations (e.g. Grinin, 
1994).
The LMC and SMC HAeBe stars show a peculiar irregular colour 
magnitude variation: they are "bluer-when-fainter" (see star ELHC2 in figure 
1). 
This reverse colour effect might be  due to the contribution of a
reflection nebula. The dust clouds that
produce the light curves are small compared to the
reflection nebula and may diminish the light of the star,
while most of the (blue) scattered light from the reflection nebula 
is unattenuated. The net result is a decrease in brightness at all
wavelengths, but slightly less in B than in V. This effect rarely
occurs in observations of Galactic HAeBe stars, (only during very 
deep
minima) because the photometric aperture is much smaller than the
projected reflection nebulae. 

We suggests that the blueing effects of LMC/SMC HAeBes is because we 
are observing the stars  at 50 kpc, and therefore not resolving their
associated
\begin{figure}
\epsfxsize=15cm
\epsfysize=8cm
\centerline{\epsffile{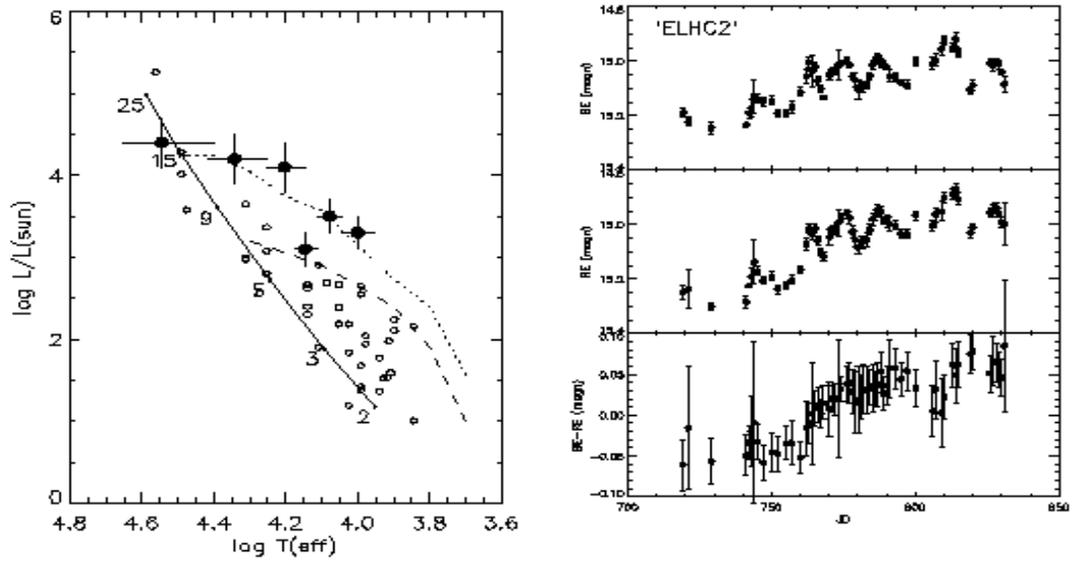}}
\caption[]{ 
{\bf Left figure}: The location of the first six LMC HAeBe stars in the 
HRD (points)  with the Galactic HAeBe stars (circles). The LMC HAeBe stars are well above
 the galactic upper limit ("birthline"). Dashed line: predicted birthline
for an accretion rate $10^{-5} M_\odot~yr^{-1}$ and dotted line for
 $10^{-4} M_\odot~yr^{-1}$ from Palla and Stahler (1993). We recall 
that 
{\bf Right figure}: The light curve and the colour curve of the LMC
 HAeBe star ELHC2.  The time is in days since Jan 1 1990. 
Notice the irregular variations and the curious 
 fact the star is "bluer-when-fainter". }
\end{figure} 
 reflecting nebulae.
For instance, the reflecting nebulae around V380 Ori about 1 arcmin
in diameter would be only 0.6 arcsec projected in the LMC. So the
scattered blue light from the nebulae will contribute significantly 
to
the EROS photometry. We have modeled this effect by scaling the 
photometric variations of
the HAeBe star  V380 Ori and its nebulae to a distance of the LMC. 
The redder when fainter behavior of V380 becomes a bluer when 
fainter behavior.
\acknowledgments
This work is based on observation held at ESO la Silla.
JPB is  grateful to IAU for financial support in attending IAU190, and
to members of the EROS and PLANET collaborations. 


%
%

%


\begin{references}
%
\reference Afonso et al. (EROS coll.), 1998, \astap, 337, L17
%
\reference Albrow M. et al. (PLANET coll.), 1998 \apj, in press
%
\reference Alcock C., et al. (MACHO coll.), 1993, Nature, 365, 621
%
\reference Alcock C., et al. (MACHO coll.), 1997, \apj,486, 697
%
\reference Alcock C., et al. (MACHO coll.), 1998, \apjlett, 499, 9
%
\reference Ansari R., et al. (EROS coll.), 1995, \astap, 299, L21
%
\reference Aubourg E., et al. (EROS coll.), 1993, Nature, 365, 623
%
%
\reference Beaulieu J.P., et al. (EROS coll.), 1995, \astap, 299, 168
%
\reference Beaulieu J.P., et al. (EROS coll.), 1995, \astap, 303, 137
%
 \reference Beaulieu J.P. \& Sasselov D.D. 1997, in  Variable stars 
and the astrophysical 
return of microlensing surveys, eds Ferlet R. \& Maillard J.P., 
editions frontieres.
%
\reference Beaulieu J.P. et al. (EROS coll.), 1996, Science 272, 995
%
\reference Beaulieu J.P. et al. (EROS coll.), 1997a, \astap, 318, L47
%
\reference Beaulieu J.P. et al. (EROS coll.), 1997b, \astap, 321, L5
%
%
\reference Beaulieu J.P., \& Sackett P.D., 1998 \aj 116, 209
%
\reference Bernasconi P. \& Maeder A., 1996, \astap, 307, 829
\reference de Laverny P., et al. (EROS coll.), 1998, \astap, 335, L93
%
\reference  Grinin V.P., 1994, in Nature and evolutionary status of
Herbig Ae/Be stars ASP conf Ser. Vol 62., ed. P.S.Th\'e.
 
\reference Grison, P., et al. (EROS coll.), 1995 \astap, 109, 445
%
\reference Hill V. \& Beaulieu J.P., 1997, in Variable stars and the 
astrophysical  return of microlensing surveys, eds Ferlet R. \& Maillard J.P., 
editions frontieres.
%
\reference Lamers H.J.G.L.M., Beaulieu J.P., de Wit W.J. 1998, 
\astap, in press
\reference Palanque-Delabrouille N, et al. (EROS-2 coll.), 1998 
\astap,  332, 1  
%
\reference Palla F. \& Stahler S.W. 1993, \apj, 418, 414 
%
\reference Renault C., et al., 1998a (EROS coll.), \astap, 324, L69
%
\reference Renault C., et al., 1998b (EROS coll.), \astap 329, 522
%
%
\reference Sasselov et al., 1997 (EROS coll.), \astap, 324, 471
%
\reference Welch D.L., et al.,  1999 (MACHO coll.), this volume

\reference Zhao H.S., 1998 \mnras, 294, 139

\reference Zaritsky D. \& Lin D., 1997, \aj 114, 254

\end{references}
\end{document}